# Surface Passivation of III-V GaAs Nanopillars by Low Frequency Plasma Deposition of Silicon Nitride for Active Nanophotonic Devices

Bejoys Jacob[1]‡, Filipe Camarneiro[1]‡, Jérôme Borme[2], Oleksandr Bondarchuk[3], Jana B. Nieder[1]**, and Bruno Romeira[1]*

[1]INL – International Iberian Nanotechnology Laboratory, Ultrafast Bio- and Nanophotonics group, Av. Mestre José Veiga s/n, 4715-330 Braga, Portugal

[2]INL – International Iberian Nanotechnology Laboratory, 2D Materials and Devices group, Av. Mestre José Veiga s/n, 4715-330 Braga, Portugal

[3]INL – International Iberian Nanotechnology Laboratory, Advanced Electron Microscopy, Imaging and Spectroscopy Facility, Av. Mestre José Veiga s/n, 4715-330 Braga, Portugal

**ABSTRACT:** Numerous efforts have been devoted to improve the electronic and optical properties of III-V compound materials via reduction of their non-radiative states, aiming at highly-efficient III-V sub-micrometer active devices and circuits. Despite many advances, the poor reproducibility and short-term passivation effect of chemical treatments such as sulfidation and nitridation requires the use of protective encapsulation methods, not only to protect the surface, but to provide electrical isolation for device manufacturing. There is still a controversial debate on which combination of chemical treatment and capping dielectric layer can best reproducibly protect the crystal surface of III-V materials, while being compatible with readily available semiconductor-foundry plasma deposition methods. This work reports on a systematic experimental study on the role of sulfide ammonium chemical treatment followed by dielectric coating (either silicon oxide or nitride) in the passivation effect of GaAs/AlGaAs nanopillars. Our results conclusively show that, under ambient conditions, the best surface passivation is achieved using ammonium sulfide followed by encapsulation with a thin layer of silicon nitride by low frequency plasma enhanced chemical deposition. Here, the sulfurized GaAs surfaces, the high level of hydrogen ions and the low frequency (380 kHz) excitation plasma that enable intense bombardment of hydrogen, all seem to provide a combined active role in the passivation mechanism of the pillars by reducing the surface states. As a result, we observe up to a 29-fold increase of the photoluminescence (PL) integrated intensity for the best samples as compared to untreated nanopillars. X-ray photoelectron spectroscopy analysis confirms the best treatments show remarkable removal of gallium and arsenic native oxides. Time-resolved micro-PL measurements display nanosecond lifetimes resulting in a record-low surface recombination velocity of ~$1.1 \times 10^4$ cm s$^{-1}$ for dry etched GaAs nanopillars. We achieve robust, stable and long-term passivated nanopillar surfaces which creates expectations for remarkable high internal quantum efficiency (IQE>0.5) in GaAs nanoscale light-emitting diodes. The enhanced performance paves the way to many other nanostructures and devices such as miniature resonators, lasers, photodetectors and solar cells opening remarkable prospects for GaAs active nanophotonic devices.

**INTRODUCTION**

A wide-range of nanoscale light sources have been reported employing III-V materials as the gain medium.[1,2] These semiconductor compound materials are crucial for the fabrication of miniaturized optical sources such as nanoscale light-emitting diodes (nanoLEDs),[3] and nanolasers,[4] of growing importance for compact photonic integrated circuits (PICs) needed in optical data communications,[5] optical computing including neuromorphic computing,[6–8] sensing and spectroscopy,[1,2,9] as well as medical diagnosis applications. Noteworthy, in the past few years remarkable developments in nanoLEDs have been made using either semiconductor III-V or III-V on silicon materials. The approach for miniaturization relies on the use of cavities such as photonic crystals,[10] metal-dielectric[11] or plasmonic,[3] thus enabling the realization of wavelength and subwavelength scale devices. Alternative nanoLED architectures also include the use of other material systems such as fin-shaped semiconductors,[12] even though extreme current densities are required in this case. These advances are creating expectations that nanoLEDs can be both efficient and fast, thereby capable of outperforming nanolasers.[13,14] However, to this date the external quantum efficiency (EQE) at room-temperature of III-V nanoLEDs remains limited to values below 1%, resulting in ultralow output powers (in the nW or even pW range),[3,10,11] which makes nanoLEDs challenging for practical optical systems. Taking the example of III-V nanopillars, and neglecting losses related with metallic structures in metal-dielectric or plasmonic nanocavites, the main reasons for the extremely low EQEs are two-fold. Firstly, coupling the light output efficiently to a nanowaveguide,[11] or a plasmonic waveguide,[3] remains a challenge when the area of the light source is reduced to the deep sub-µm.[2] Secondly, at these small scales non-radiative effects in III-V materials, specifically surface-related properties, become more important as the surface-to-volume ratio increases substantially. In this work, we devote our attention to the role of the non-radiative effects in the performance of III-V gallium arsenide (GaAs) light-emitting subwavelength devices.

Among the wide range of III-V materials available for active nanophotonic devices, the GaAs/AlGaAs is one of the most studied and a key compound material for photonics,[4,15–18] providing optical emission and absorption in a wide range of wavelengths spanning from the visible to near-infrared (NIR). GaAs has recently been notable in many photonic applications such as 3D sensing using GaAs-based lasers, NIR-LEDs and visible red-orange-yellow LEDs for displays. However, the surface of GaAs-based materials and their interfaces with dielectrics tend to host large densities of electronically active defects (or dangling bonds).[19] As a result, at ambient conditions, an oxide layer is formed on the surface of the GaAs (e.g. $Ga_2O_3$ and $As_2O_3$), which leads to charge trapping.[20] Importantly, when semiconductors are nanostructured, namely using top-down dry etching, the plasma reactive etching process can induce additional surface damages,[21] such as surface roughness due to ion bombardment, surface contamination due to polymer deposition, or surface stoichiometry change due to preferential etching. Overall, this results in charge trapping effects (i.e. non-radiative active centers), leading in the case of GaAs nanoscale LEDs to extremely short lifetimes (sub-100 ps),[16] and ultralow efficiencies.[11,22]

A wide range of methods have been reported for passivating GaAs surfaces of micro- and nanoscale structures and devices.[10,17,18] One technologically challenging and expensive method is the epitaxial growth of a high bandgap layer on the GaAs surface.[23–26,27,28] The high bandgap layer reduces the surface trap density since it prevents carriers in GaAs from accessing the surface states and thus reduces the photoluminescence decay rate. On a second approach, chemical passivation, including nitridation[29,30] and sulfidation[20,31–34] by wet chemistry are inexpensive and widely used methods. Sulfidation for example has proven to be effective in removing the native oxides and elemental arsenic from the surface by creating an S termination on the semiconductor surface.[20,31,32] Still, this termination tends to be unstable when exposed to air or water and the passivation procedure is strongly dependent on the chemical composition, light and temperature conditions which makes it difficult to achieve reproducible results. Nitridation in bulk GaAs samples has been recently shown to be more robust and resistive to air over about 100 hours.[30] In GaAs optical wavelength-sized optical structures (disk resonators),[35] wet nitridation revealed to increase substantially their optical quality factor. But in all scenarios, protective layers are still needed, not only to prevent the sulfide or nitride layer's degradation (due to oxidation or other environmental effects), but also to provide electrical isolation for optoelectronic device manufacturing.

Several deposition methods can produce dielectric films such as silicon oxide ($SiO_2$), silicon nitride ($Si_3N_4$), and alumina ($Al_2O_3$), with excellent properties, including atomic layer deposition (ALD) and plasma enhanced chemical vapor deposition (PECVD). Interestingly, a number of studies suggest that the surface passivation can be highly sensitive to the structure and composition of the semiconductor-dielectric interface, and the interface formation process may depend on hydrogen content, stoichiometry and density of the ALD- and PECVD-fabricated films, and also on subsequent temperature treatments.[36] As a result, the protective layers can not only prevent degradation of the surface, but also play an active role on the passivation effect.[37] Recently, it has been reported that not only the type of protective film but also the frequency of the plasma deposition (specifically lower RF excitation) can play an important role on the passivation of *n*-type GaAs electronic devices,[38] due to the ionic bombardment inherent to the low frequency plasma. Nevertheless, its impact in the optical properties of GaAs-based semiconductors and nanostructures has been overlooked, and to our knowledge, totally unexplored.

In this study, we report on an experimental investigation to identify which combination of chemical passivation and dielectric protective film layer could best reproducibly passivate and protect the crystal surface of III-V materials while being compatible with readily available semiconductor-foundry plasma deposition methods. Specifically, we present a systematic experimental study that investigates the role of the sulfide ammonium chemical treatment followed by various dielectric coatings ($SiO_x$ or $Si_xN_y$) by either low frequency or high frequency PECVD in the surface passivation effect of unintentionally doped GaAs/AlGaAs compound semiconductor nanopillars. Our results conclusively show that, under ambient conditions, the best passivated surfaces of sub-µm GaAs/AlGaAs deeply etched nanopillars are achieved using a combination of ammonium sulfide followed by encapsulation with a thin film layer of $Si_3N_4$ (~80 nm) deposited by low frequency PECVD at 300 °C. For this surface treatment, a remarkable 29-fold enhancement of the photoluminescence (PL) intensity is achieved for the best samples as compared to untreated nanopillars. We observe a robust, stable and long-term (>10 months) passivation effect for nanopillars ranging from 200 nm to 1 µm. The quality of the passivation treatment can be quantified by the minimum amount of surface defects formed by the native oxides on the GaAs surface and this has been analyzed by X-ray photoelectron spectroscopy. The measurements show successful removal of gallium and arsenic native oxides for the best treatment using sulfurization of GaAs pillars immediately followed by HF-PECVD $Si_xN_y$ deposition, which is in line with the PL measurements. Time-resolved PL measurements reveal that the lifetimes of the best passivated nanopillars can reach a lifetime of ~1 ns, leading to estimations of a record-low surface recombination velocity of ~$1.1 \times 10^4$ cm s$^{-1}$ for dry etched GaAs-based nanopillars. This value compares to some of the best passivated core-shell GaAs/AlGaAs nanowires ($1.7 \times 10^3$ cm s$^{-1}$ to $1.1 \times 10^4$ cm s$^{-1}$) of similar width dimensions. However, our method uses a conventional semiconductor-foundry industrial-ready PECVD deposition method instead of challenging and expensive epitaxial growth methods.[27] These results demonstrate the impact of the surface passivation on the internal quantum efficiency (IQE) of passivated GaAs-based light-emitting pillars which could reach an IQE>0.5. Our results pave the way for III-V GaAs active nanophotonic devices such as nanoLEDs and nanolasers operating at room-temperature with large efficiencies, and other relevant sub-µm structures such as nano-waveguides and miniature resonators.

**EXPERIMENTAL SECTION**

**Fabrication of GaAs/AlGaAs nanopillars.** A systematic experimental study was performed to investigate the passivation effect on AlGaAs/GaAs/AlGaAs nanopillars. The semiconductor layer stack, Fig. 1(a), was composed from top to bottom, by 150 nm of AlGaAs (30% Al), 52 nm of a GaAs-based compound material consisting of a GaAs (20 nm)/AlAs (3nm)/GaAs (6 nm)/AlAs (3nm)/GaAs (20 nm) double barrier quantum well (DBQW) nanostructure, 150 nm of AlGaAs (30% Al), and 300 nm of GaAs, all not intentionally doped, and grown by molecular beam epitaxy on a GaAs substrate. The selection of the GaAs-DBQW nanostructure is motivated by its quantum resonant tunneling phenomenon for applications in electrically-pumped nonlinear LED sources, of relevance for neuromorphic nanophotonic computing.[8] In this work, we are mainly interested on the passivation effect on the GaAs/AlGaAs layer stack to achieve efficient light emission. The fabrication of the nanopillars involved nanopatterning via electron beam lithography using a Vistec 5200 ES 100 kV tool. The pillars were dry etched until $H$~0.54 µm depth ($H$ is the height of the nanopillar) using inductively coupled plasma (ICP) in an SPTS ICP machine (the nanofabrication description can be found in Supporting Information S1).

Figure 1(b) displays the scanning electron microscope (SEM) image of representative fabricated semiconductor pillars. The samples contained pillars with dimensions ranging from 200 nm to 1 µm width, $d$, organized in arrays spaced by at least 10 µm so that the emission could be collected and analyzed individually from each single pillar. On the same sample (not shown) micropillars with dimensions ranging from 3 µm – 8 µm width were also fabricated. Figure 1(c) shows the SEM picture of a $d$~400 nm wide circular nanopillar, and Figure 1(d) shows an example of a micropillar (1 µm width). As a result of the dry-etching step the nanopillars typically displayed sloped sidewall features with an angle ~17°.

**Surface passivation treatments.** A set of identical samples containing micro- and nanopillar arrays were fabricated as discussed previously. After fabrication of the nanopillars, the surface passivation entailed the following six main treatment procedures (Table 1). Treatment #1 was a sulfur treatment only consisting of a 20% ammonium sulfide solution that was further diluted [$H_2O$:$(NH_4)_2S$ (10:1)], where samples were dipped for 5 minutes at 65 °C under dark conditions. In treatment #2, the samples were submerged in ammonium sulfide solution, similarly as described in treatment #1. Then, immediately after the sulfur a thin capping layer of $SiO_x$ was deposited by high frequency (RF excitation source of 13.56 MHz) PECVD. In treatment #3, the thin capping layer of $SiO_x$ was deposited immediately after the sulfur using low frequency (RF excitation source of 380 kHz) PECVD. Here the RF plasma was tuned well below the ion transit frequency (estimated ~2 MHz). In treatment #4, immediately after the sulfurization a thin capping layer of $Si_xN_y$ was deposited by high frequency (13.56 MHz) PECVD. In treatment #5 instead of high frequency PECVD, the $Si_xN_y$ deposition was performed by low frequency (RF excitation source 380 kHz) PECVD. Lastly, in treatment #6, the samples were coated with a thin capping layer of $Si_xN_y$ deposited also by low frequency (380 kHz) PECVD but without employing the ammonium sulfide solution pre-treatment. All film depositions were performed with a substrate temperature of 300 °C. A complete description of treatments #1–#6 can be found in Supporting Information S2. For the purpose of comparing both passivated and unpassivated pillars under the same fabrication processing conditions, for each passivation treatment (Table 1), an unpassivated sample of pillars was simultaneously fabricated and left uncoated without any sulfurization treatment.

**Steady-state and time-resolved micro-photoluminescence.** The photoluminescence of fabricated nanopillars and the effect of the respective surface passivation treatment (Table 1) was measured using a micro-photoluminescence (µPL) setup consisting of a Witec Alpha 300R confocal microscopy system fiber-coupled to a UHTS300 spectrometer coupled to an Andor Peltier cooled CCD detector. In our measurements we have used a continuous-wave laser at 532 nm wavelength (2.33 eV energy) under low pumping conditions. The optical emission from the pillars was collected using a 100× air objective with a high numerical aperture (Supporting Information S3). The PL decay was measured in a time correlated single photon counting (TCSPC) experimental setup described in the Supporting Information S4. In short, the output of a pulsed laser diode at 561 nm (2.21 eV), with a pulse with a full-width half-maximum (FWHM) of ~80 ps and a repetition rate of 50 MHz, was used for excitation of the micro- and nanopillars. The pillars were optically pumped using a 100× high numerical aperture oil immersion objective.

**Energy-dispersive X-ray spectroscopy (EDS) and X-ray photoelectron spectroscopy (XPS).** The quality of the passivation treatments can be quantified by the amount of surface defects formed by the gallium and arsenic native oxides, Ga-O ($Ga_2O_3$) and As-O ($As^{3+}$ and $As^{5+}$), respectively. The analysis of these native oxides on the GaAs surface was performed using EDS and XPS. Initial surface characterization studies of fabricated samples employed EDS analysis with a scanning electron microscope (FEI NovaNanoSEM 650), equipped with an EDS system (detailed description in Supporting Information S6). The XPS spectra was collected using an ESCALAB 250Xi system in UHV (< $10^{-9}$ Torr). A monochromatic Al-Kα source (1486.6 eV) was used to analyze an area of 650 μm×650 μm in prepared samples. Since XPS spectra can effectively be collected using only thicknesses within 10 nm from the surface, for the XPS measurements samples were prepared with deposited dielectric coatings with a thicknesses of ~4 nm, instead of ~80 nm (a detailed description can be found in Supporting Information S7).

**RESULTS AND DISCUSSION**

**Steady-state PL spectroscopy.** The pillars' PL was characterized in dependence of the applied chemical pre-treatments, the dielectric coatings, and in dependence of the plasma frequency of the coating-deposition, as summarized in Table 1. The table shows the value of PL improvements as compared with unpassivated samples taken for the best samples for the case of a 400 nm wide nanopillar. As a summary, the chemical treatment with sulfurization showed only minor PL improvements (~1.3-fold). The passivation with sulfurization followed by $SiO_x$ coatings (deposited by either by LF- or HF-PECVD) showed an intermediate performance with PL improvements ranging from 2 to 4-fold. Lastly, the passivation treatments using $Si_xN_y$ coatings (either by LF- or HF-PECVD) displayed the best performance with PL improvements for the best samples ranging from 5 to 29-fold. Particularly, the results indicate the best improvements (up to 29-fold) are achieved for nanopillars encapsulated with a layer of $Si_xN_y$ deposited by low frequency PECVD. Next, the PL results for each treatment #1 to #6 are analyzed and discussed in detail.

Figure 2(a) displays the typical μPL spectra of a representative 400 nm wide nanopillar for treatment #1 (sulfurization only) showing the typical luminescence for both passivated and unpassivated cases. For the tests realized neither the sulfurization realized at room-temperature (results not shown) nor at 65 °C (Fig. 2(a)) revealed meaningful improvements. This can be expected since it is known that the reproducibility of sulfurization passivation treatments is strongly dependent on the temperature, light conditions, pH and composition of the solution making it difficult to achieve reproducible results. In Figure 2(b) it is shown the μPL spectrum for treatment #2 (sulfurization followed by $SiO_x$ coating). Interestingly, here a 4-fold increase in the PL integrated intensity is observed. We note however this result is much lower than the improvements shown for other III-V materials (e.g. InGaAs) using a similar procedure.[37] An identical passivation treatment but using $SiO_x$ deposition by LF-PECVD instead (treatment #3, PL not shown) did not reveal substantial PL improvements (~2-fold) as compared with unpassivated samples. This follows similar studies which consistently report that the passivation of GaAs shows the best results when $Si_xN_y$ coating materials are employed, as previously shown in field effect transistors,[39] or terahertz emitter devices.[40]. This is attributed not only to the excellent source of hydrogen for further passivating the residual interface defect states that can be obtained when the $Si_xN_y$ is deposited by PECVD,[39] but also to the fact that $Si_xN_y$ films coated in an initial clean surface can additionally participate directly in the formation of interfacial bonding at the GaAs surface in such a way as to reduce the density of defect sites.[41]

Indeed, our tests reveal that all the treatments using $Si_xN_y$ layers show the best results as compared to $SiO_x$. For example, treatment #4 ($Si_xN_y$ by high frequency PECVD) shows a PL improvement of up to 6-fold as compared to $SiO_x$. Remarkably, when the $Si_xN_y$ film is deposited by LF-PECVD immediately after the sulfurization (treatment #5), up to a 29-fold PL intensity increase (measured in an identical 400 nm wide nanopillar) was achieved (red trace of Fig. 3(a)) as compared to the unpassivated sample (black trace). Figure 3(b) shows an histogram of the integrated spectra summarizing the results from treatments #2 (blue), #5 (red) and untreated (black) pillars as a function of the pillar width. PL improvements were achieved for pillars ranging from 1 µm down to 400 nm showing a PL enhancement ranging from 22-fold to 29-fold, respectively, as compared with untreated pillars. PL enhancements are observed also in the bulk surface region of the etched GaAs material and for micropillar sized pillars (>1 µm) indicating an impressive passivation effect in either sub-µm/µm etched structures, or bulk materials (S5, Fig. S3). Noteworthy, Fig. 3(c) presents measurements of the nanopillar sample shown in Fig. 3(a) but recorded after 10 months (the samples were stored with regulated conditions at a temperature of 2o°C and a humidity of 40%). A similar PL (red trace) enhancement of ~28-fold is achieved. The results indicate a stable and long-term passivation effect. The PL is compared with the same untreated sample shown in Fig. 3(a) that was protected in month zero with $SiO_x$ to avoid further oxidation (we note this procedure did not affect the initial PL measured in Fig. 3(a), black trace).

Finally, Fig. 3(d) displays the µPL spectra of an identical 400 nm pillar for the unpassivated case and for treatment #6, that is, using $Si_xN_y$ coating by low frequency PECVD only, and without employing any chemical pre-treatment. Improvements of the PL (~5-fold) were achieved indicating the low frequency plasma indeed plays a role in the passivation effect. As a result, this dry-only single-step passivation method using low frequency plasma shows a unique potential to be used in industrial environments for highly reproducible, simple and cost efficient passivation methods. Lastly, we note that when comparing the spectra from unpassivated and passivated samples, the emission wavelength peak for the passivated samples typically ranges from ~854nm to 858 nm. This emission is attributed to the central 52 nm DBQW GaAs active region and is consistent with the expected electron to heavy/light-hole bandgap transitions from the 20 nm GaAs QW layers surrounding the AlAs barriers. The unpassivated or poorly passivated samples show emission mainly peaking at ~865 nm corresponding to emission from the bottom GaAs region ~190 nm, and therefore close to the band-edge emission expected for a GaAs bulk material (~872 nm) - Supporting Information S5, Fig. S3. This suggests that after successful passivation a pronounced emission enhancement effect is achieved particularly for the GaAs-DBQW active material.

**EDS and XPS analysis.** The quality of the passivation treatments can be further quantified by the amount of surface defects formed by the native oxides on the GaAs surface [here Ga-O ($Ga_2O_3$) and As-O ($As^{3+}$ and $As^{5+}$)]. In this section we focus our analysis on the removal of these native oxides by the treatments employing $Si_xN_y$ layers that showed the best PL improvements. For initial surface characterization studies of PECVD $Si_xN_y$ treatments we used EDS in a scanning electron microscope system (see Supporting Information S6). For the pillars measured (pillar width 200 nm – 1 µm) traces of oxygen were not identified, Fig. S4, indicating a good passivation of PECVD $Si_xN_y$ treatments. We note is challenging to quantify in the EDS analysis of our SEM system the presence of native oxides below 1 atomic percentage (at %), in particular light atoms. As a result, to quantify and compare the removal of gallium and arsenic oxides in the various treatments we focused our attention on samples measured by XPS.

Figure 4 shows the Ga 3d XPS spectra comparison for an untreated sample, Fig. 4(a), and for samples using various $Si_xN_y$-based surface treatments, Figs. 4(b)-(d). First we analyze the passivation using LF-PECVD $Si_xN_y$ without any sulfurization pre-treatment. In the unpassivated case, Fig. 4(a), we observe a high energy shoulder which is less pronounced for the LF-PECVD $Si_xN_y$, Fig. S4(b). This indicates suppression of the Ga native oxide (Ga-O) peak (binding energy ~20 eV, blue trace), showing the LF-PECVD without pre-treatment provides already an impact on the removal of gallium oxides. Noteworthy, this effect is noticeable even in the case of a thin deposited layer (~4 nm). We note this thin layer was a requirement in our experiments to be able to perform the XPS analysis.

Next we compare the LF-PECVD $Si_xN_y$ treatment with the HF-PECVD $Si_xN_y$, Figs. 4(c) and 4(d), respectively. Clearly in both cases the GaAs peak (binding energy ~19.2 eV) is the prominent peak whereas Ga native oxides (Ga-O) are insignificant. This shows the success of combining the ammonium sulfide and $Si_xN_y$ coatings for the removal of native oxides. Analyzing in more detail both cases, we observe a broader and larger Ga-O peak for the HF-PECVD $Si_xN_y$ coated sample, panel (d), as compared to the LF-PECVD Si$_x$N$_y$ coated sample, panel (c). This indicates a better performance of the LF-PECVD Si$_x$N$_y$. These results are confirmed in Table S1 (Supporting Information S7) which summarizes the ratio of the atomic percentage of Ga-O to GaAs. A remarkable low at % ratio (~0.1) is achieved for LF-PECVD Si$_x$N$_y$ coated sample (as compared with an at % ratio of ~0.3 for the HF-PECVD Si$_x$N$_y$) which indicates the least presence of Ga-O, in line with the improvements measured in PL. A similar analysis of the arsenic oxide (As-O) peaks was performed for the As 3d XPS spectra (see Supporting Information S7). As discussed in the Supporting Information, Fig. S6 and Table S2, a complete suppression of native As-O oxides is achieved using the ammonium sulfide combined either with LF-PECVD or HF-PECVD, which is in line with the trend observed in our PL measurements showing the best PL improvements for these treatments.

Following the PL results and the XPS analysis, we attribute the success of our best treatments as the combined effect of three crucial factors: firstly, sulfide ammonium with immediate coating enables to remove native oxides and protects the surface from further re-oxidation; secondly, additional native oxide removal using PECVD coating of Si$_x$N$_y$ is achieved by the high level of hydrogen injection provided by plasma dissociation of SiH$_4$ and NH$_3$, making H$^+$ the most concentrated ion in the plasma (such a mechanism has also previously been argued to be responsible for improved passivation[39]), and thirdly, the low-frequency PECVD increases ion bombardment as H$^+$ ions that are able to follow the excitation RF signal and reach the substrate surface after plasma ignition which is able to further remove the presence of surface states. We note the kinetic energy of ions, particularly hydrogen, gets significantly higher under the ion transit low frequency (typically below 2 MHz), resulting in ions that are able to follow the RF excitation which then reach immediately the surface after plasma ignition. For example, recent work on metal-insulator-semiconductor capacitors fabricated by depositing Si$_x$N$_y$ on *n*-doped GaAs at a frequency of 90 kHz,[38] reports a low density of surface states (in that case ~$10^{11}$ cm$^{-2}$ eV$^{-1}$) due to the intense ionic bombardment related to the low frequency RF excitation.

Noteworthy, the fact that deposition of other dielectrics (here $SiO_x$) known to contain significant hydrogen was not found to result in similar passivation improvements strongly suggests that more than simply hydrogenation is occurring. Therefore, it is possible that the $Si_xN_y$ film additionally participates directly in the formation of interfacial bonding at the GaAs surface, either supplementing or substituting the existing S-terminated bonds, in such a way as to reduce the density of defect sites. However, we note the exact mechanisms of the passivation effect under RF excitation in the properties of the $Si_xN_y$/GaAs interface could be further thoroughly investigated. For example, the energy of $H^+$ ions that reach the sample can increase the surface temperature and stimulate surface diffusion,[42] which can promote chemical reconstruction leading to thermodynamically stable films. In this case, further methods such as transmission electron microscopy can be used to analyze the impact of the LF-PECVD method on the structural and morphological changes occurring on the surface of the passivated GaAs.

**Time-resolved PL spectroscopy.** To investigate the carrier dynamics in GaAs/AlGaAs pillar structures, we performed time-resolved photoluminescence spectroscopy (TRPL) measurements using a time-correlated single-photon counting (TCSPC) setup (Supporting Information S5). Due to the expected extremely short lifetimes (<<100 ps), specifically for the smaller size unpassivated pillars,[16] and given the limited time resolution of our fastest detectors (~50 ps) – Supporting Information S5, Fig. S1 – we start our analysis by comparing first the micropillar devices (≥3 µm) where the measured lifetimes are well above this limit. In Fig. 5(a) the measured decay curves are shown for micropillars ($d$=3 µm) from an unpassivated (black dots trace) and two passivated samples (treatment #2 and #5, blue and red dot traces, respectively). The TRPL decay curves are fitted using a single exponential decay function to obtain the values of the carrier recombination lifetime. The results show an extremely short lifetime <150 ps for the unpassivated pillar and a lifetime ~375 ps for the case of the passivated pillar using sulfurization followed by $SiO_x$ coating. These results are in agreement with previous measurements using identical GaAs pillar structures and similar passivation methods.[43] Noticeably, for the case of the best passivated sample using low frequency PECVD deposition of $Si_xN_y$, Fig. 5(a)(right, red dot trace), the lifetime increases substantially to a value >1 ns.

In order to quantify the surface recombination velocity, *S*, of the measured pillars, we assume that under the low excitation conditions employed in the experiments, the surface-related non-radiative recombination rate scales as $4S/d$,[37] so that *S* can be estimated directly from the size-dependent carrier lifetimes in the low injection regime:

$$\frac{1}{\tau_{PL}} = \frac{1}{\tau_{bulk}} + \frac{1}{\tau_{SR}} = \frac{1}{\tau_b} + \frac{4S}{d} \approx \frac{4S}{d} \qquad (1)$$

where $\tau_{bulk}$ is the carrier lifetime in the bulk material. For nanoscale devices generally $\tau_b \gg \tau_{SR}$, the bulk contribution can be neglected and $\tau_{PL}^{-1} \approx 4S/d$. This allows us to directly convert the measured lifetime to the surface recombination lifetime. Figure 5(b) shows the inverse carrier lifetime estimated from the TRPL measurements versus the inverse pillar width (4/*d*), before and after the passivation treatments presented in Fig. 5(a). The corresponding linear fit (dashed black curves) of the experimental data, Fig. 5(b), allows us to estimate a surface recombination of $5.54 \times 10^5$ cm s$^{-1}$ for unpassivated samples, and of $2.66 \times 10^4$ cm s$^{-1}$, that is, a 20-fold improvement for the LF-PECVD $Si_xN_y$ treated samples, which is in line with the trend observed in the PL measurements.

We note in our results the decay curves for unpassivated or poorly passivated samples are exponential and therefore the surface dominated recombination still remains valid. However, particularly for the best passivated samples (e.g. Fig. 5(a)(right)), the PL can also exhibit a nonexponential decay, even when very low pumping conditions are employed. This has been reported not only for GaAs semiconductors (e.g. nanowires[27]) but also for InGaAs[37] and InGaAsP[36] nanostructures. Typically, this nonexponential behavior is attributed to the radiative recombination of mobile charges which is a bimolecular process ($\propto BN^2$, where B is the bimolecular recombination coefficient and *N* is the photoexcited carrier density). This leads to the assumption that the initial decay curve is dominated by radiative recombination and then decays to the surface dominated decay rate towards longer photon arrival times (specifically for highly passivated samples). Effectively, the decay curves can be modelled taking into account both surface and bimolecular recombination.[36] However, there is still a debate on the exact phenomena that can contribute to the non-exponential behavior, which can be strongly dependent, among other factors, on the semiconductor material under study.[44] For example, other recombination mechanisms such as trap-assisted non-radiative charge recombination, formed for example by defects, impurities and dangling bonds, or inhomogeneous distribution of trap energy can contribute to this behavior.[30] The main goal of this paper is not to study all mechanisms of charge recombination and therefore for simplicity of analysis and to better compare our results with literature, here the data presented is quantified using a single exponential fit since the weight of second component is rather small and therefore has a negligible contribution to the calculated lifetimes.

Next we analyze the TRPL decay curves of sub-µm pillars for the best sample in treatment #5 ($Si_xN_y$ coating by LF-PECVD) that shows a remarkable 29-fold increase of PL, Fig. 3(a). Figure 6(a) shows the decay curves for a few representative pillars with 400 nm, 600 nm and 800 nm pillar width. The PL lifetimes increase from 0.92 ns to 0.98 ns and 1 ns, respectively. We note in all other measurements of poorly passivated samples (not shown), the measured lifetimes were well below the instrument response function of our setup (Supporting Information S4, Fig. S1), and therefore their lifetimes are expected to be extremely short (<<100 ps). Applying $\tau_{PL}^{-1} \approx 4S/d$, as described previously, the calculated surface velocity ranges from $1.1 \times 10^4$ cm s$^{-1}$ to $2 \times 10^4$ cm s$^{-1}$. These results indicate a record-low surface velocity for dry etched GaAs-based nanopillars which is comparable to the best core-shell passivated GaAs nanowires[27] of comparable width dimensions ($S$ in the range of $1.7 \times 10^3$ cm s$^{-1}$ to $1.1 \times 10^4$ cm s$^{-1}$). However, these methods require challenging and expensive epitaxial growth methods. Our results show substantially improvements as compared to other methods such as doping of GaAs nanowires[16] for enhanced radiative efficiency ($S \sim 2.18 \times 10^6$ cm s$^{-1}$).

We note that by increasing the pump conditions from low pumping with a pump fluence of ~1.5 µJ cm$^{-2}$, Fig. 6(a), to a mid-pump fluence of ~40 µJ cm$^{-2}$, Fig. 6(b), the nanopillars can exhibit lifetimes longer than 1 ns, and therefore lifetimes comparable with micropillar sized structures. Figure 6(c) presents an overview of all nanopillars by showing the corresponding PL lifetime images of the nanopillars ranging from 200 to 1000 nm under the same pumping conditions as presented in Fig. 6(b). A clear contrast to the background of the sample is achieved indicating that effectively the measured lifetimes are a result of the successful passivation of the pillars' surface. The lifetimes range between 0.74 ns for the smallest nanopillars (200 nm) up to ~0.95 ns for the 1 µm pillar size. The lifetimes exhibit slightly shorter values than the ones presented in the single histogram results, Fig. 6(b), which is related to lifetime binning used in the image analysis (as exemplified in Supporting Information S4, Fig. S2).

**Internal quantum efficiency.** To further characterize the effect of the large reduction of the surface recombination velocity on the internal quantum efficiency (IQE) of pillars, we have calculated the IQE, which is the ratio of radiative emission rate ($\tau_r^{-1} = Bn^2$) to the sum of non-radiative and radiative emission rate ($\tau_{nr}^{-1} + \tau_r^{-1}$), for the case of a 400 nm pillar width for both the best passivated ($S=1.1\times10^4$ cm s$^{-1}$) and unpassivated ($S=5.54\times10^6$ cm s$^{-1}$) scenarios. In this analysis, Auger recombination, $C$,[45] was also considered, so that the non-radiative term reads $\tau_{nr}^{-1} = (4S/d)n + Cn^3$. As displayed in Fig. 7(a), in the low concentration regime (carrier density of $3\times10^{17}$ cm$^{-3}$), IQE values of ~0.04 are obtained for the passivated case while this value drops substantially to an IQE of $9.2\times10^{-4}$ in the case of the unpassivated sample. Remarkably, for a larger carrier density concentration ($10^{19}$ cm$^{-3}$), where nanoLEDs are expected to operate,[46] a high value of IQE=0.54 is calculated for the 400 nm sized pillar, limited only by Auger recombination. This is a 20-fold improvement as compared to the unpassivated nanopillar (IQE~0.028). This analysis illustrates the strong role of the non-radiative effects on the low efficiency reported in III-V nanolight sources.[3,10,11] As shown in Fig. 7(b), the IQE for the Si$_x$N$_y$ passivation case remains high ($\geq$0.1) for all analyzed pillar sizes (0.2 µm–1 µm) and for the carrier density values considered ($10^{18}$ cm$^{-3}$ and $10^{19}$ cm$^{-3}$). We note these results indicate a best case scenario and do not take into account other factors that can play a role in the efficiency of nanopillar devices, such as carrier injection efficiency. The results reported here, combined with the enhancement of the light extraction efficiency for identical sub-µm GaAs/AlGaAs pillars reported elsewhere,[43] could lead to substantial improvements of the external quantum efficiency of nanostructures such as nanopillars or nanorods, for example when integrated in photonic crystal cavities, optical resonators, or coupled to waveguides, etc. Therefore, our findings are of key importance for the miniaturization of GaAs optical components and devices.

**CONCLUSIONS**

We have successfully passivated the surface of GaAs/AlGaAs nanopillars using a combination of ammonium sulfide chemical treatment followed by encapsulation with silicon nitride, a widely used dielectric, by low frequency plasma deposition. We demonstrate up to a 29-fold increase of the photoluminescence integrated intensity at room-temperature for the best passivated nanopillar samples as compared to unpassivated nanopillars. This leads to estimations of a low surface velocity of $\sim 1.1 \times 10^4$ cm s$^{-1}$ for dry etched GaAs-based nanopillars. The wide-range of tests and analysis performed, including XPS analysis to investigate the amount of surface defects, confirm that the best passivation treatment is a combination of three crucial factors: firstly, sulfurization of GaAs surfaces with immediate coating enables to remove native oxides without further re-oxidation. Importantly, sulfurization prepares the initial surface for the coating material; secondly additional native oxide removal using PECVD coating of $Si_xN_y$ is achieved by the high level of hydrogen injection; and thirdly the low frequency (380 kHz) plasma enables intense ionic bombardment of H$^+$ ionic species as a result of the RF excitation, playing an active role in the passivation of nanopillars by further removing the presence of surface states. We note that according to previous studies,[38] in principle the low frequency effect should be observed for a wide-range of frequencies of the plasma as long as the selected frequency is under the ion transit low frequency (typically below 2 MHz). Since our PECVD system uses two fixed RF power generators at 380 kHz and 13.56 MHz, it was unpractical to implement further studies with varying frequencies. Taking advantage of this passivation method, the low frequency plasma shows a unique potential to be used in industrial environments as a highly reproducible and cost-effective passivation method needed for the exponentially growing miniaturized

GaAs devices and applications, namely electrically-pumped nanopillar LEDs. Since GaAs-based devices typically require post rapid annealing temperature treatments for electrical contacts annealing, we identify that further studies on the stability of the passivation would be relevant. In fact, several studies show that passivation treatments benefit substantially from post-annealing,[36] which could improve the results achieved here. Importantly, the passivation method based on low frequency PECVD can potentially be extended to other III-V materials covering additional wavelengths and be exploited for a wide range of high-performance room-temperature nano-optoelectronic active devices such as nanoLEDs, nanolasers, nanophotodetectors needed for energy-efficient emerging photonic integrated circuit technologies, with applications in neuromorphic or quantum photonic computation, bioimaging, information and communication technologies and internet of things, or improved performance of nanostructured solar cells.

## AUTHOR INFORMATION

**Corresponding Author**

*bruno.romeira@inl.int, **jana.nieder@inl.int

**Author Contributions**

The manuscript was written through contributions of all authors. All authors have given approval to the final version of the manuscript. ‡These authors contributed equally

**Funding Sources**

European Commission, H2020-FET-OPEN Project "ChipAI" under Grant Agreement 828841. CCDR-N (NORTE-01-0145-FEDER-000019).

**Notes**

The authors declare no competing financial interest.

ACKNOWLEDGMENT

This work was supported by the European Commission through the H2020-FET-OPEN Project "ChipAI" under Grant Agreement 828841. The authors acknowledge the discussion on passivation treatments of III-V semiconductors with Ekaterina Malysheva and Victor Calzadilla, Eindhoven University of Technology. We acknowledge the Micro and Nanofabrication Facility and the Nanophotonics & Bioimaging Facility, and the Advanced Electron Microscopy Facility at INL.

Table of Contents artwork

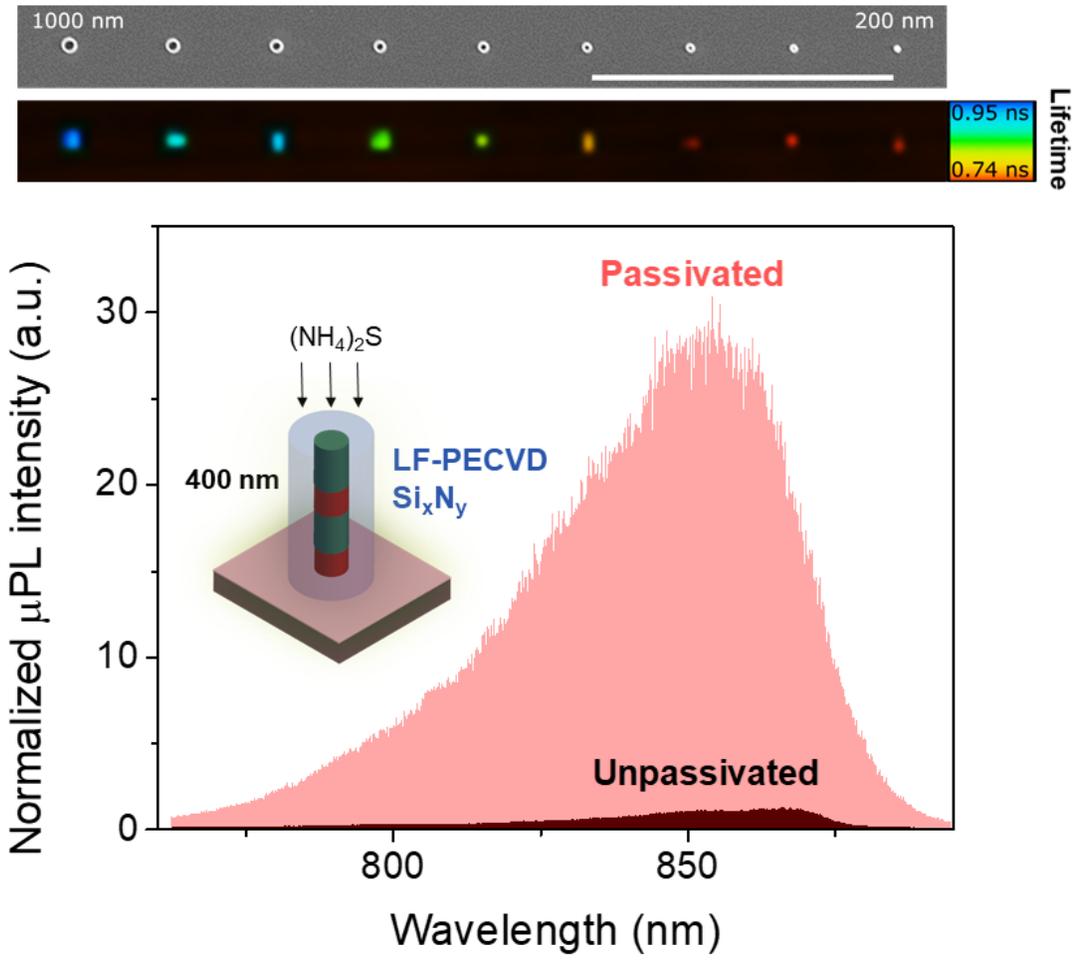

TABLES

**Table 1.** Summary of the main surface treatments showing the treatment description, value of PL improvement as compared with unpassivated sample and schematic of the passivated bonds at the surface of the GaAs materials for the various treatments. The value of the PL integrated intensity improvement was taken for the best samples for the case of a 400 nm wide nanopillar.

| Surface treatment # | Treatment description | PL of best sample | Passivated bonds |
|---|---|---|---|
| #1. Chemical treatment: $(NH_4)_2S$ | I. Sample submerged in ammonium sulfide solution, $(NH_4)_2S$, at 65 °C during 5 minutes. | 1.3-fold | GaAs — S, O, S, S |
| #2. $(NH_4)_2S$ + $SiO_x$ coating (HF-PECVD) | I. Sample submerged in ammonium sulfide solution, $(NH_4)_2S$, at 65 °C during 5 minutes. II. Deposition of 80 nm thick silicon oxide, $SiO_x$, capping by HF-PECVD (13.56 MHz) at 300 °C. | 4-fold | GaAs — S, S, S, S / $SiO_x$ |
| #3. $(NH_4)_2S$ + $SiO_x$ coating (LF-PECVD) | I. Sample submerged in ammonium sulfide solution, $(NH_4)_2S$, at 65 °C during 5 minutes. II. Deposition of 80 nm thick silicon oxide, $SiO_x$, capping by LF-PECVD (380 kHz) at 300 °C. | 2-fold | GaAs — S, S, S, S / $SiO_x$ |
| #4. $(NH_4)_2S$ + $Si_xN_y$ (HF-PECVD) | I. Sample submerged in ammonium sulfide solution, $(NH_4)_2S$, at 65 °C during 5 minutes. II. Deposition of 80 nm thick silicon nitride, $Si_xN_y$, capping either by HF-PECVD (13.56 MHz) at 300 °C. | 6-fold | GaAs — S, S, S, S / $Si_xN_y$ |
| #5. $(NH_4)_2S$ + $Si_xN_y$ (LF-PECVD) | I. Sample submerged in ammonium sulfide solution, $(NH_4)_2S$, at 65 °C during 5 minutes. II. Deposition of 80 nm thick silicon nitride, $Si_xN_y$, capping either by LF-PECVD (380 kHz) at 300 °C. | 29-fold | GaAs — S, S, S, S / $Si_xN_y$ |
| #6. $Si_xN_y$ coating only | I. Deposition of 80 nm thick silicon nitride, $Si_xN_y$, capping by LF-PECVD (380 kHz) at 300 °C. | 5-fold | GaAs — O, O, O, O / $Si_xN_y$ |

FIGURES

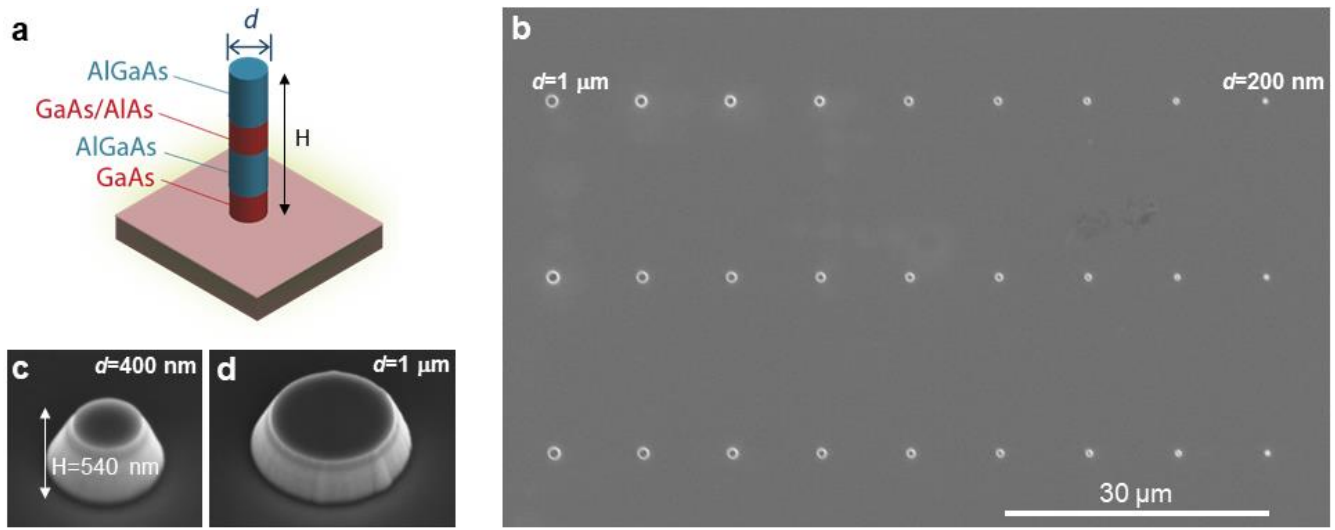

**Figure 1.** Scheme and fabricated III-V semiconductor GaAs/AlGaAs nanopillars. (a) Schematic of a GaAs/AlGaAs nanopillar. SEM images of (b) GaAs/AlGaAs pillar array; (c) a 400 nm wide nanopillar; and (d) a 1 μm wide micropillar.

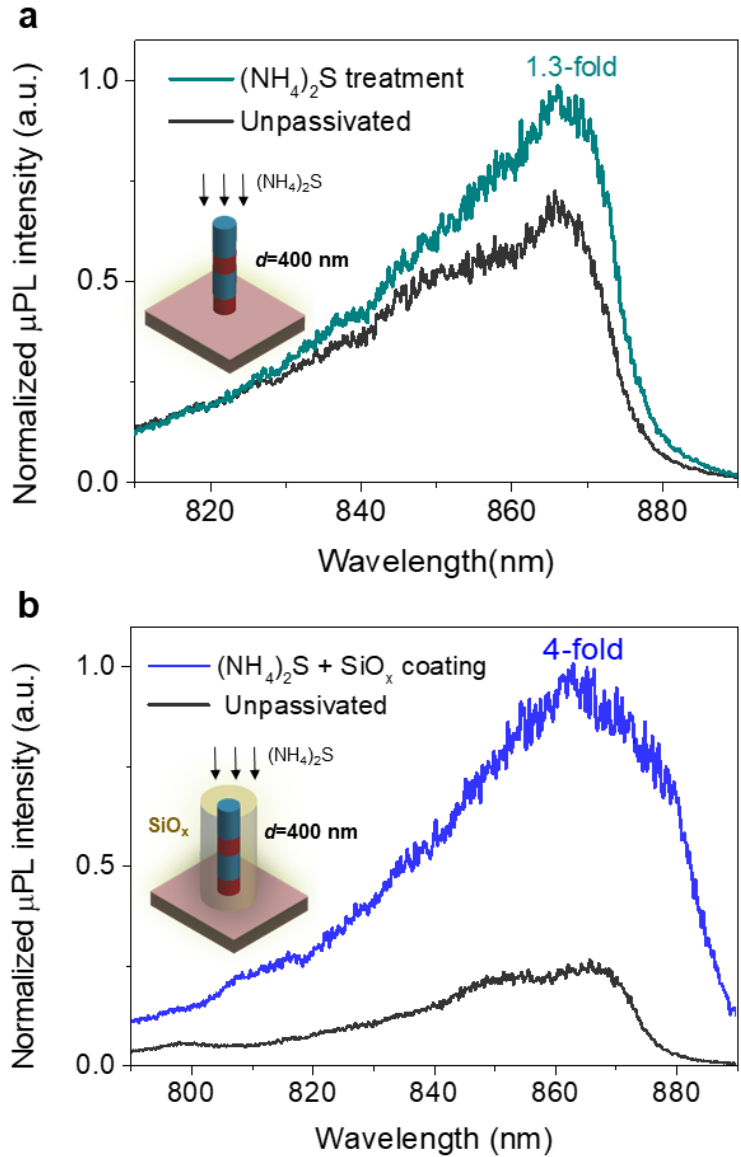

**Figure 2.** Continuous-wave photoluminescence measurement results at room-temperature displaying a typical μ-PL spectra from a single nanopillar with around 400 nm width for (a) unpassivated and sulfur passivation treatment passivation steps, and (b) unpassivated and ammonium sulfide followed by SiO$_x$ coating deposited by HF-PECVD passivation treatment passivation steps.

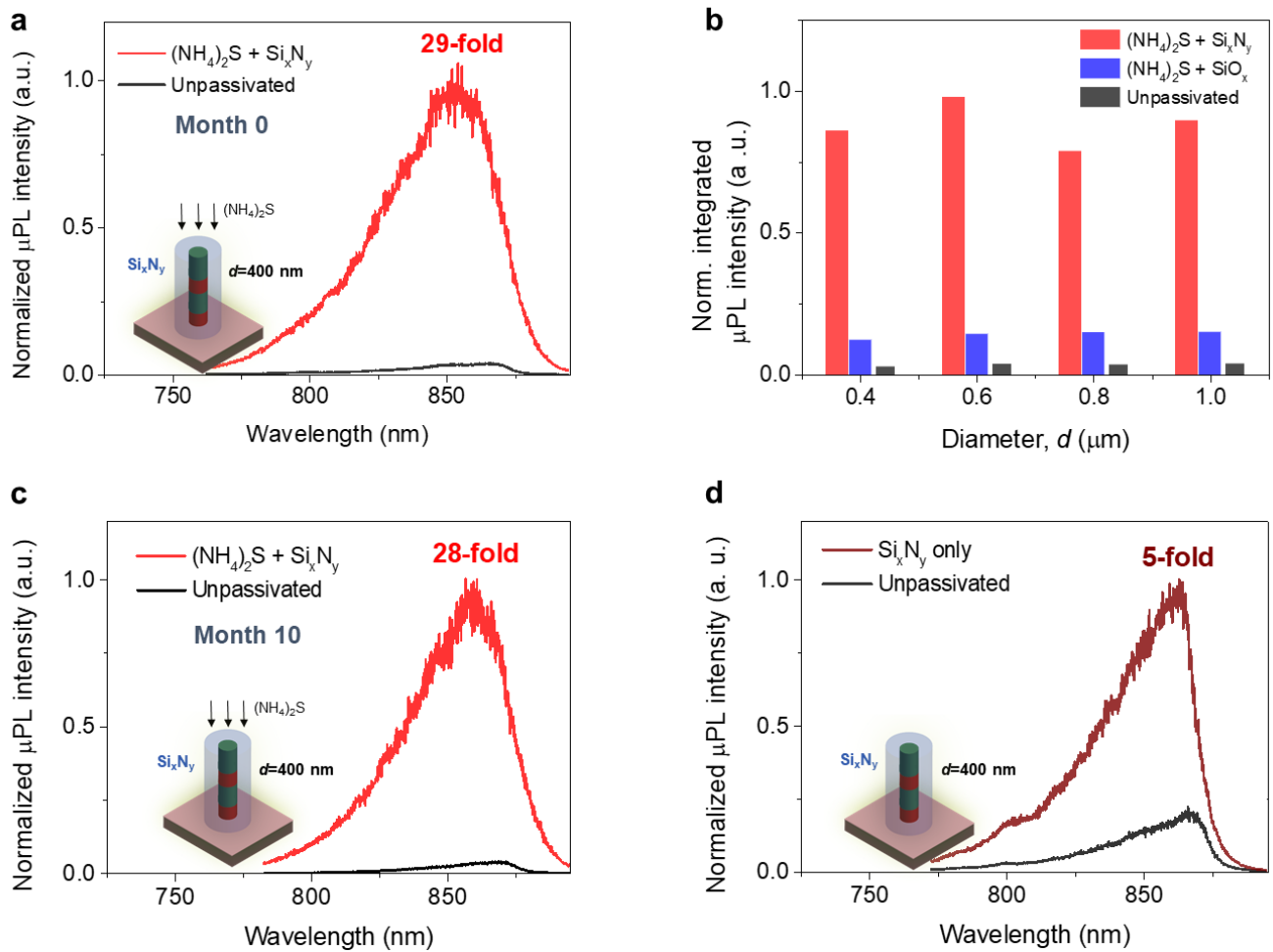

**Figure 3.** Experimental continuous-wave photoluminescence measurement results at room-temperature. (a) Photoluminescence results displaying a typical μ-PL spectra from a single nanopillar with around 400 nm width for an unpassivated sample (black curve) and a sample treated ammonium sulfide followed by $Si_xN_y$ coating deposited by LF-PECVD (red curve). Inset is shown a schematic of the best treatment displaying a pillar coated with $Si_xN_y$ dielectric. (b) Normalized intensity as a function of the nanopillar width for all 3 untreated and passivation treatment cases for pillar diameters ranging from 400 nm to 1 μm. (c) Repeated PL measurements for the same samples shown in panel (a) after 10 months. (d) Photoluminescence results displaying the μ-PL spectra for unpassivated samples and samples using $Si_xN_y$ coating by low frequency PECVD, without chemical pretreatment.

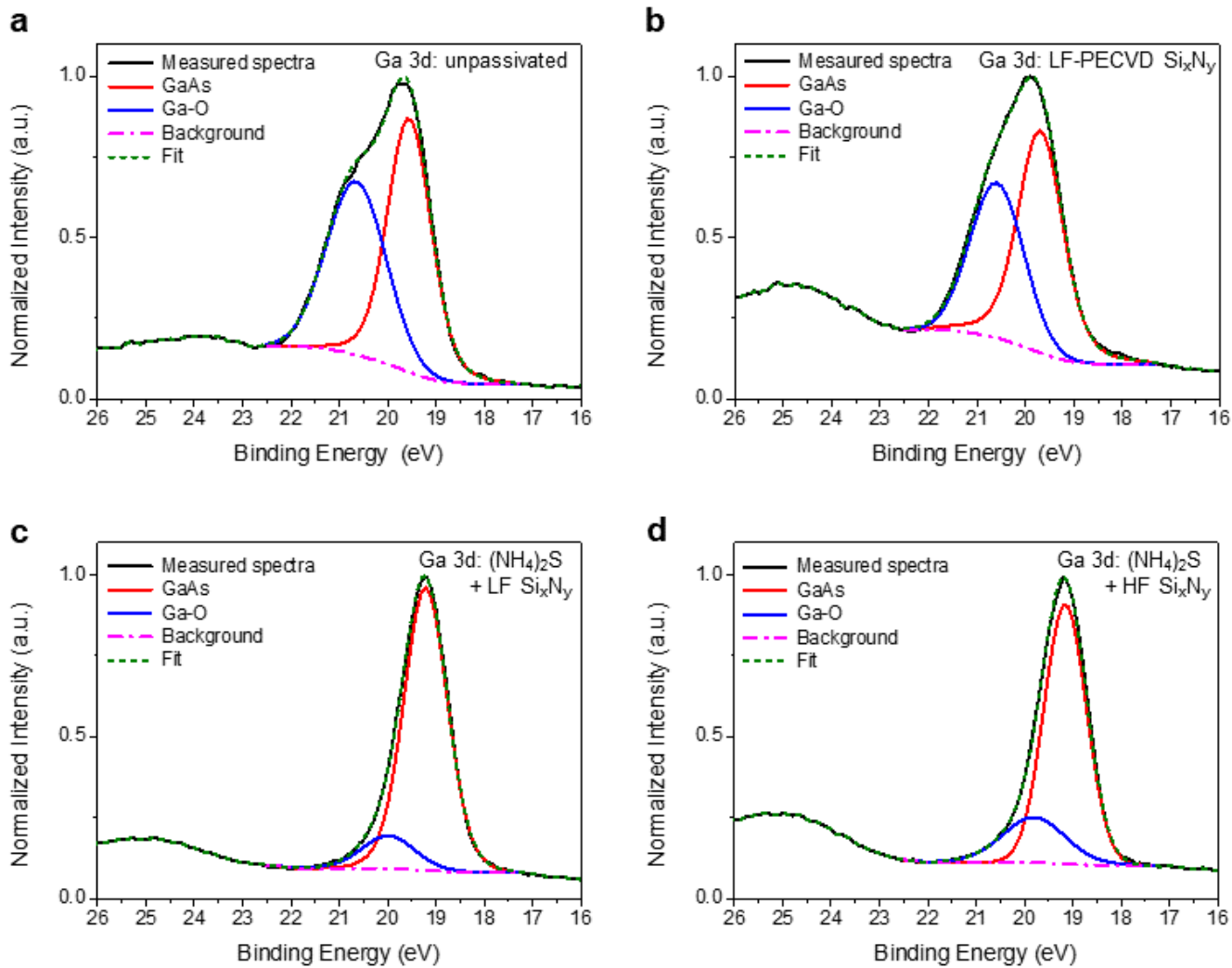

**Figure 4.** XPS spectra of Ga 3d. (a) Unpassivated sample. (b) Sample coated using LF-PECVD $Si_xN_y$. (c) Sample using ammonium sulfide treatment followed by LF-PECVD $Si_xN_y$ coating. (d) Sample using ammonium sulfide treatment followed by HF-PECVD $Si_xN_y$ coating.

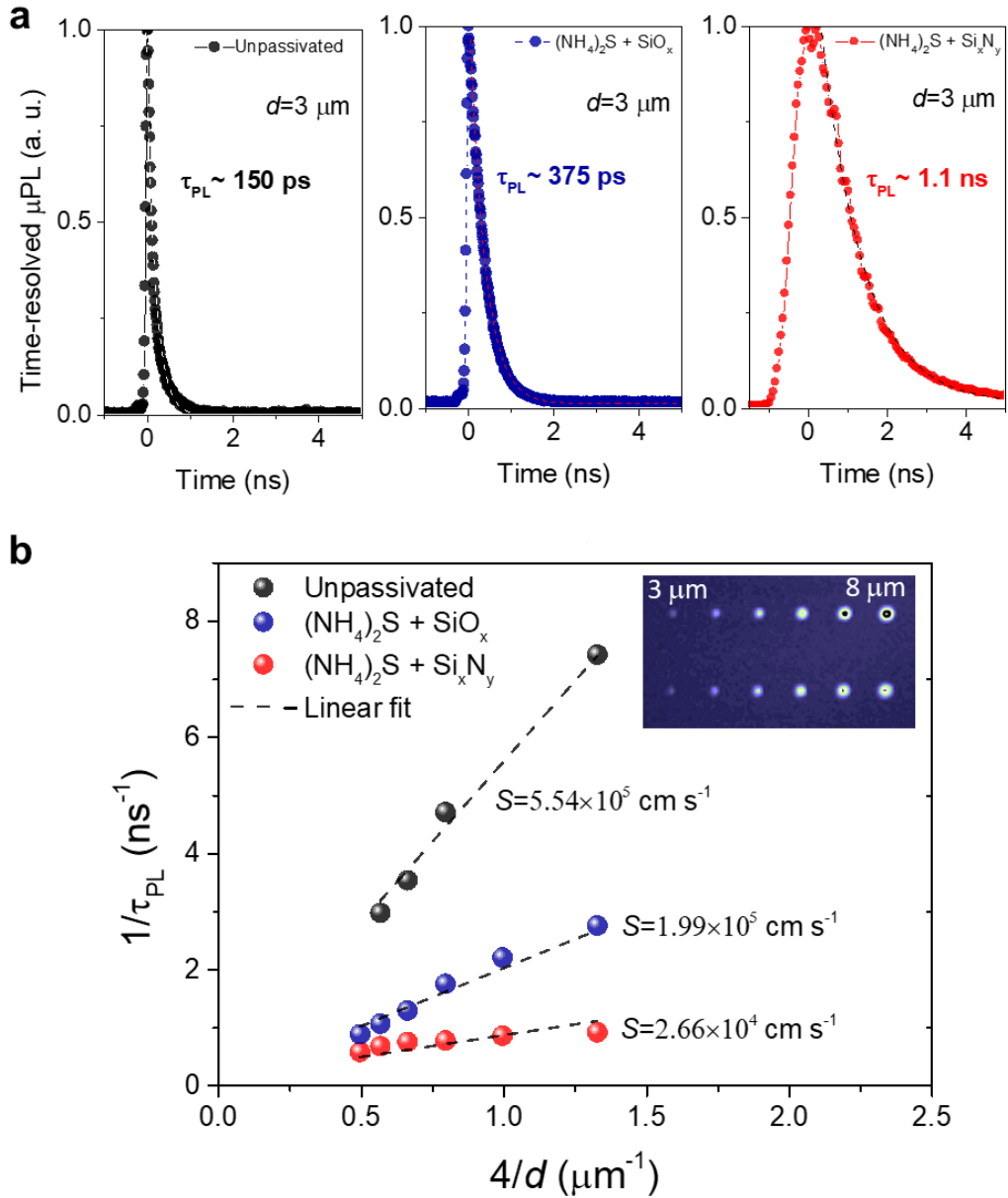

**Figure 5.** Experimental TRPL decay curves of micropillars measured at room temperature. (a) TRPL decay curves measured for a representative 3 μm wide micropillar using the treatments: (left): unpassivated (grey dots); (center): passivated coated pillars with $SiO_x$ coating (treatment #2, purple dots); (right): passivated coated pillars with $Si_xN_y$ coating by low frequency PECVD (treatment #5, red dots). The TRPL decay time is quantified using 1/e method (dashed lines). (b) Inverse carrier lifetime, $1/\tau_{PL}$, estimated from the TRPL measurements versus the inverse pillar width, $4/d$, before passivation (grey dots), using $SiO_x$ coating (blue dots), and using $Si_xN_y$ coating (red circles). Also shown is the corresponding linear fit (dashed black curves) of the experimental data for estimation of the corresponding surface recombination

velocity, $S$. Inset is shown a microscope intensity image of the measured micropillar array for the $SiO_x$ passivated sample with pillar widths ranging from 3 μm and 8 μm.

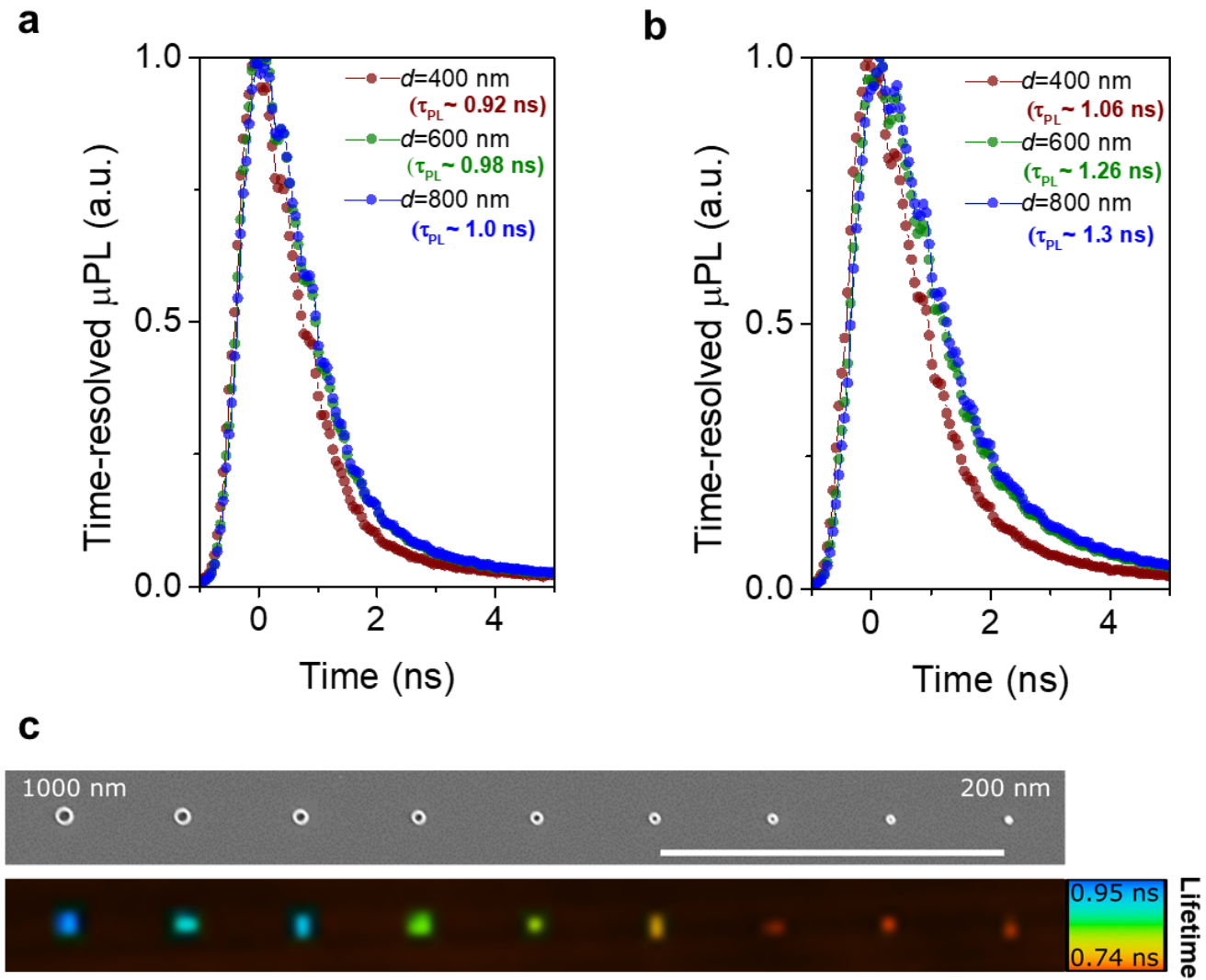

**Figure 6.** Experimental TRPL decay curves of nanopillars measured at room temperature for the best passivated samples using $Si_xN_y$ coating by low frequency PECVD (treatment #5). (a) TRPL decay curves under low pumping conditions (pump fluence ~1.5 μJ cm$^{-2}$). (b) TRPL decay curves under moderate pumping conditions (pump fluence ~40 μJ cm$^{-2}$) showing nanosecond lifetimes for nanopillars. The TRPL decay time in all plots is quantified using 1/e method. (c) FLIM image (S4) as a function of pillar diameter for pillars ranging from 1 μm to 200 nm and under the same pumping conditions as presented in panel (b). On top of the map a SEM image of the nanopillars is shown (scale bar is 30 μm).

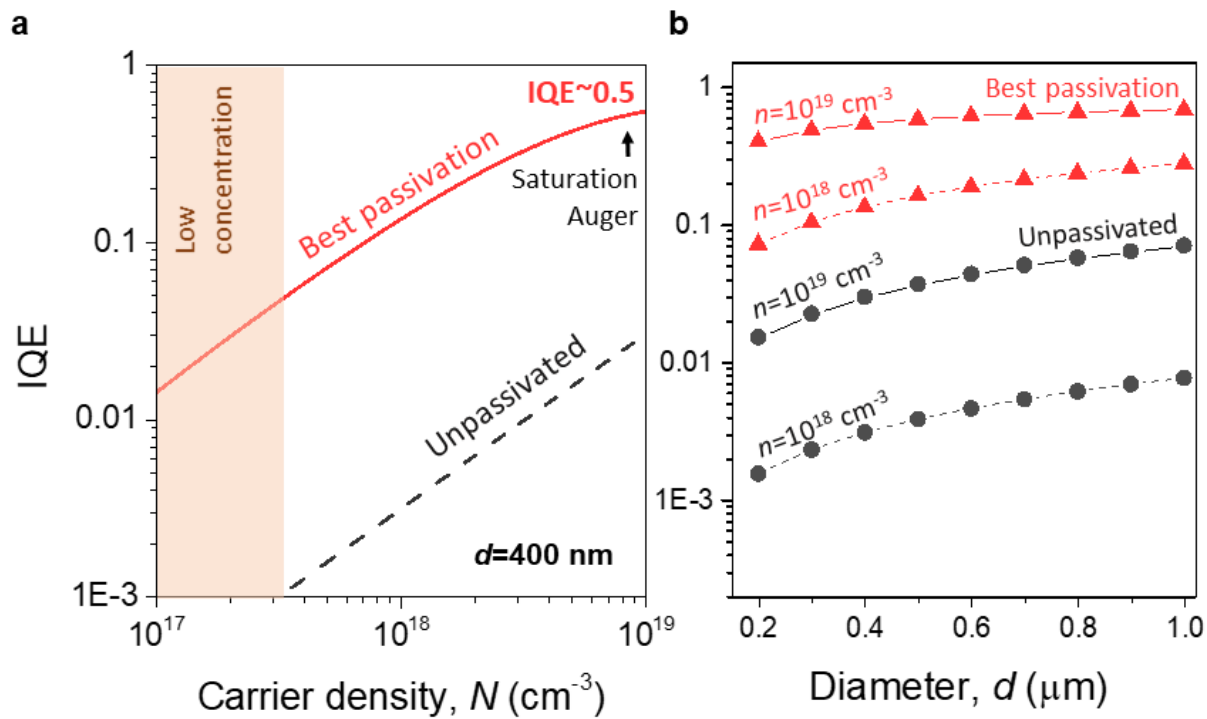

**Figure 7.** Theoretical internal quantum efficiency of nanopillars. (a) Estimated IQE for unpassivated and passivated pillars using the case of a representative pillar width of $d$=400 nm. (b) IQE versus pillar diameter taken at the carrier density of $10^{18}$ cm$^{-3}$ and $10^{19}$ cm$^{-3}$ for both unpassivated and passivated pillars. All plots assumed Auger recombination effect using typical values for GaAs material, $C$=3.5×10$^{-30}$ cm$^6$ s$^{-1}$.

# Supporting Information

## Surface Passivation of III-V GaAs Nanopillars by Low Frequency Plasma Deposition of Silicon Nitride for Active Nanophotonic Devices

Bejoys Jacob[1]‡, Filipe Camarneiro[1]‡, Jérôme Borme[2], Oleksandr Bondarchuk[3], Jana B. Nieder[1]**, and Bruno Romeira[1]*

[1]INL – International Iberian Nanotechnology Laboratory, Ultrafast Bio- and Nanophotonics group, Av. Mestre José Veiga s/n, 4715-330 Braga, Portugal

[2]INL – International Iberian Nanotechnology Laboratory, 2D Materials and Devices group, Av. Mestre José Veiga s/n, 4715-330 Braga, Portugal

[3]INL – International Iberian Nanotechnology Laboratory, Advanced Electron Microscopy, Imaging and Spectroscopy Facility, Av. Mestre José Veiga s/n, 4715-330 Braga, Portugal

Corresponding Authors
*bruno.romeira@inl.int, **jana.nieder@inl.int

‡ These authors contributed equally

## S1. Fabrication of the GaAs/AlGaAs micro- and nanopillars

The fabrication used samples cleaned first using acetone and isopropanol (IPA) to remove the photoresist layer followed by 13 minutes oxygen plasma at 230 W to remove any organic material and followed by a deoxidation treatment of 2 minutes in diluted $NH_4OH:H_2O$ (1:10). The fabrication of the micro- and nanopillars involved an electron beam lithography (EBL) step using a 5200 ES 100 kV tool from Vistec, where the pillars were patterned using a negative e-beam resist (ARN7520.18) of 500 nm thickness with a 200 nm $SiO_x$ hard mask deposited by plasma enhanced chemical vapor deposition [PECVD (model CVD MPX, a machine from SPTS)] at 300°C. Before the resist deposition a treatment of hexamethyldisilazane (HMDS), a primer deposited in an oven at 150 ºC was used to favor uniformity and adhesion in the resist deposition. After exposure of the resist using EBL, baking on a hot plate at 85 °C for 60 s is performed, followed by a 120 s of development using developer AR 300.47 (TMAH) diluted 4:1 in water. The sample is cleaned with water and the developer solution refreshed every 30 s. The sample is finally spin dried. The next step involved transferring the pattern from the resist to the hard mask using a module for reactive ion etching (APS, a machine from SPTS) (etch rate of ~594 nm/min). Once the pattern was transferred, the remaining resist was removed by 13 minutes oxygen plasma at 230 W followed by a deoxidation step of 2 minutes in diluted $NH_4OH:H_2O$ (1:10). The sample was then rinsed in ultra-pure water (UPW) and dried with an $N_2$ pistol. The following step was to etch the pillars by dry etching with inductively coupled plasma (ICP, a machine from SPTS) using a mixture of Ar and $BCl_3$ chemistry at 40°C to deeply etch the pillars until ~0.54 µm depth, followed by a cleaning step were the samples were rinsed in UPW and dried with an $N_2$ pistol. After, the remaining $SiO_x$ hard mask was etched with hydrofluoric (HF) acid in a vapor etcher tool (Primaxx uEtch, a machine from SPTS) using anhydrous HF vapor at 13% during 600s, followed by one cycle of 13 minutes oxygen plasma treatment at 230 W to clean the surface of the pillars. The last step was a deoxidation step using a solution of $NH_4OH:H_2O$ (1:10) for 2 min. After, the samples were rinsed in UPW and dried with an $N_2$ pistol.

## S2. Surface passivation treatments

After fabrication of the nanopillars, the surface passivation procedures consisted in the following six main treatments.

**Surface treatment #1: $(NH_4)_2S$ chemical treatment.** The samples, previously treated diluted in $NH_4OH:H_2O$ (1:10) were after immediately treated with a sulfur treatment consisting of ammonium sulfide as a passivation agent, where the samples were submerged in a diluted solution of $H_2O:(NH_4)_2S$ (1:10) for 5 minutes at 65 °C under dark conditions, prepared using an ammonium sulfide solution, 20% in $H_2O$. The samples were then dried with an $N_2$ pistol, without rinsing in between.

**Surface treatment #2: $(NH_4)_2S$ + $SiO_x$ coating by HF-PECVD.** The samples were submerged in ammonium sulfide solution as described in treatment 1. For the dielectric coating step, immediately after the sulfur treatment (less than 5 minutes) a thin capping layer of $SiO_x$ of 100 nm was deposited by high-frequency (13.56 MHz) PECVD (deposition time of 126s), using 1420:10:392 sccm $N_2O:SiH_4:N_2$ as precursor gases at a pressure of 900 mTorr and power of 30 W. The deposition of non-stoichiometric $SiO_x$ was done at 300°C covering the walls and the top of the pillars.

**Surface treatment #3: $(NH_4)_2S$ + $SiO_x$ coating by LF-PECVD.** In this treatment the samples were submerged in ammonium sulfide solution as described in treatment 1, immediately followed by a thin capping layer of $SiO_x$ deposited by low-frequency (380 kHz) PECVD. The precursor gases used here consists of using 1420:12:392 sccm $N_2O:SiH_4:N_2$ as precursor gases at a pressure of 900 mTorr and power of 60 W. The deposition of non-stoichiometric $SiO_x$ was done at 300°C covering the walls and the top of the pillars.

**Surface treatment #4: $(NH_4)_2S$ + $Si_xN_y$ coating by HF-PECVD.** Immediately after submerged in ammonium sulfide the samples were coated by a layer of $Si_xN_y$ deposited by high-frequency (15.56 MHz) PECVD at 300°C. The precursor gases used here consisted of 40:55:1960 sccm $SiH_4$:$NH_3$:$N_2$ at a pressure of 900 mTorr and power of 30 W.

**Surface treatment #5: $(NH_4)_2S$ + $Si_xN_y$ coating by LF-PECVD.** The samples were submerged immediately in ammonium sulfide and then coated by a thin layer of $Si_xN_y$ of 81 nm deposited by low-frequency (380 kHz) PECVD (deposition time of 120 s). The precursor gases used here consists of 40:20:1960 sccm $SiH_4$:$NH_3$:$N_2$ at a pressure of 550 mTorr and power of 60 W. The deposition of non-stoichiometric $Si_xN_y$ was done at 300°C.

**Surface treatment #6: $Si_xN_y$ coating only by LF-PECVD.** In this treatment, the samples were not submerged in ammonium sulfide solution previously to dielectric coating. Instead, the samples previously treated diluted in $NH_4OH$:$H_2O$ (1:10) were after immediately coated with a thin capping layer of $Si_xN_y$ deposited by low-frequency (380 kHz) PECVD, using the same dielectric deposition conditions as described in surface treatment #4.

## S3. Micro-photoluminescence (µPL) characterization

The emission spectral intensity of micro- and nanopillars from the GaAs/AlGaAs layer stack semiconductor material was collected using a micro-photoluminescence microscope integrated with a spectrometer covering the visible and near-infrared region of the spectrum. It consists of a confocal Raman system in PL mode (WITec Alpha300M+, a tool from Witec Ulm) equipped with a 100× air high numerical aperture objective (NA=0.9). We have used a continuous-wave laser at 532 nm excitation wavelength (power level <100 µW) to pump the micro-and nanopillars. The collected light was filtered with a 532 nm bandpass filter and focused to a multimode fiber. The fiber-coupled light is then sent to a UHTS300 spectrometer (with 600 lines/mm grating) and then coupled to an Andor Peltier cooled CCD detector.

## S4. Time-resolved micro-PL

In the time-resolved PL measurements, excitation from a pulsed laser diode (~561 nm)(BDL-561-SMY, Becker & Hickl) was used with a pulse width ~80 ps, a repetition rate of 50 MHz and a pump fluence of ~1.5 $\mu J/cm^2$. The laser pulses are guided into a custom-built microscope based on an inverted microscope design by steering silver galvo-scanner mirrors, and expanded by a set of scan and tube lenses (SL50-CLS2 and TTL200MP, Thorlabs). The sample scanning is done via the aforementioned galvo-scanning mirrors changing the laser angle at the objective back aperture, while the sample positioning and fine focus are done via a manual XY micrometer stage and a nanometer Z-piezo stage (Nano-Z100-N, MadCityLabs). The light emitted from the pillars was collected by an oil immersion 100× high-numerical aperture objective (PFO 100x 1.3NA, Nikon). The collected light was guided to a single-photon counting avalanche photodetector (APD) (QD800c-fQ, Roithner Lasertechnik) to measure the temporal decay. For unpassivated samples showing extremely short lifetimes, see Fig. 1(a), a fastest APD (PD50CTD, MPD) was used, see respective instrument response functions (IRFs) in Fig. S1. Prior to detection, a long pass spectral filter (~561 nm) was used to select the signal of interest and filter out unwanted background signals. The APD was connected to a correlation card (TCSPC 150N, Becker & Hickl) controller. This controller correlates the photon arrival times at the APD (start signal) with the electric laser pulse arrival times (stop signal) in order to measure the luminescence decay. A histogram of these arrival times is then constructed corresponding to the time-dependent output intensity of the optically pumped pillars. The produced lifetime data was analyzed via the SPCImage software (Becker & Hickl) and via Origin using single fit algorithms. Even though some decay curves exhibit double exponential behavior, the weight of second component is rather small and therefore has a negligible contribution to the calculated lifetimes. Figure 5(b) in the paper shows a SEM image and a FLIM image of the nanopillars ranging from 200 to 1000 nm. The lifetime analysis shows a clear size dependency of the nanopillars lifetime. The lifetime values exhibit relatively shorter values than the ones presented in the single histogram results due to lifetime binning used in the image analysis. This binning takes into account the nearest neighboring pixels to calculate the lifetime of each pixel, increasing the number of counts per decay curve and improving the signal to noise ratio. The shorter lifetime values can be explained by the heterogeneous lifetime distribution inside the pillars, as exemplified for the case of micropillars (see Fig. S2), where the center of the pillars exhibit longer lifetimes than the border.

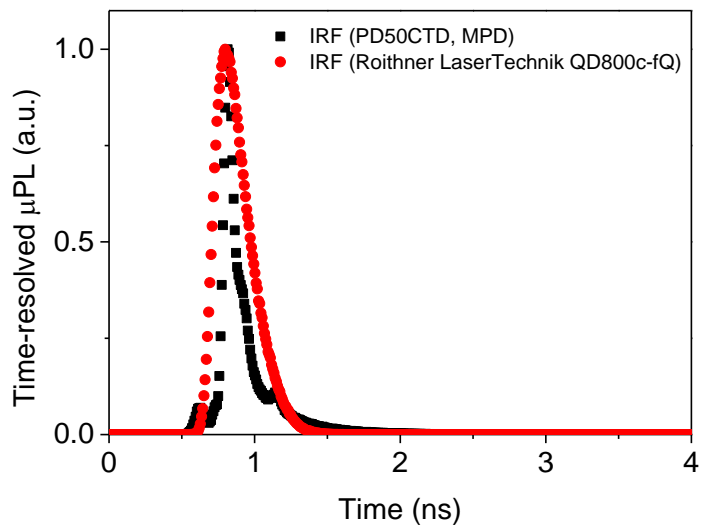

**Figure S1:** Instrument response functions of both QD800c-fQ (Roithner Lasertechnik), and PD50CTD (MPD) detectors.

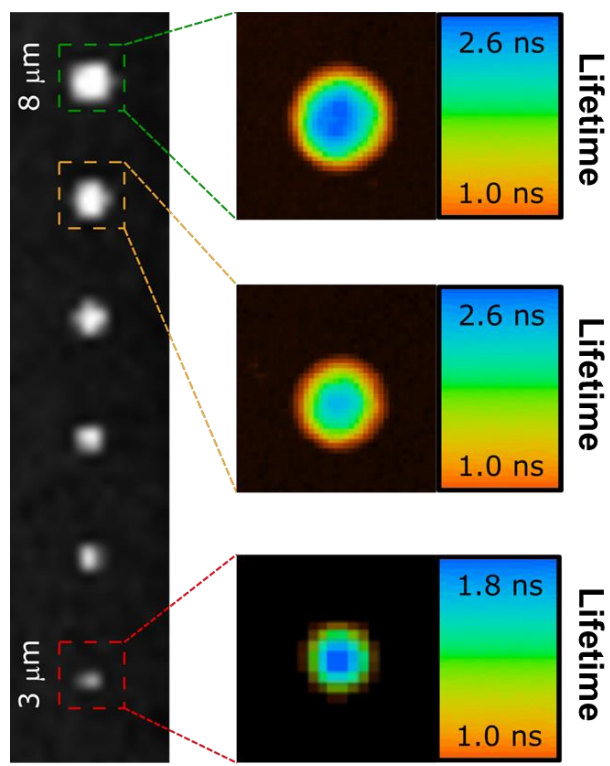

**Figure S2:** Intensity (left) and FLIM images (right) of micropillars ranging from 8 to 2 μm, taken with a 561 nm ps-laser. FLIM scans show a heterogeneous distribution of the lifetime within the micropillars. The center of the pillars exhibit longer lifetimes in comparison with the borders.

**S5. PL emission of bulk GaAs surrounding the etched pillars**

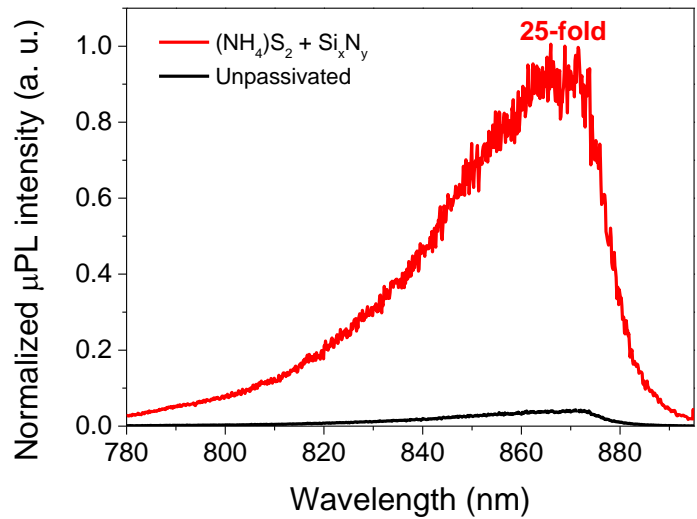

**Figure S3:** Steady state photoluminescence spectra measured at room-temperature displaying a typical μ-PL spectra from the etched bulk GaAs material surrounding the etched pillars. The spectra show the band-edge emission at ~872 nm, typical of a bulk GaAs material with a clear effect of the best passivation treatment displaying a 25-fold enhancement of the integrated PL emission. The results compare well with the typical improvements observed in nanopillars.

## S6. Energy-dispersive X-ray spectroscopy (EDS)

*Experiment*

For initial surface characterization studies using energy-dispersive X-ray spectroscopy (EDS), we used a scanning electron microscope (SEM), (FEI NovaNanoSEM 650), equipped with an EDS system (Oxford x-act). The system was operated using a voltage of 5 kV. The identification of spectral lines was performed using INCA software.

*Results and discussion*

Figure S4 shows the EDS analysis of measured pillars (pillar width ranging from 200 nm – 1 μm) for the best treatments shown in PL measurements employing $Si_xN_y$ coating layers. The results in both panels (a) and (b) do not show traces of oxygen indicating good passivation treatments. We note however the EDS analysis in our SEM system is challenging to quantify native oxides below one atomic percentage (at %), in particular for light atoms. Further, for reproducible comparison of results a good reference sample in the same percentage range as the expected change of composition is typically required. As a result, to quantify the removal of gallium and arsenic oxides we focused our attention on samples measured by X-ray photoelectron spectroscopy (XPS), see next section S7. We note in Fig. S4(a) we see traces of adventitious carbon. We attribute this to the fact that measurements were realized in samples after more than 12 months of the passivation being performed.

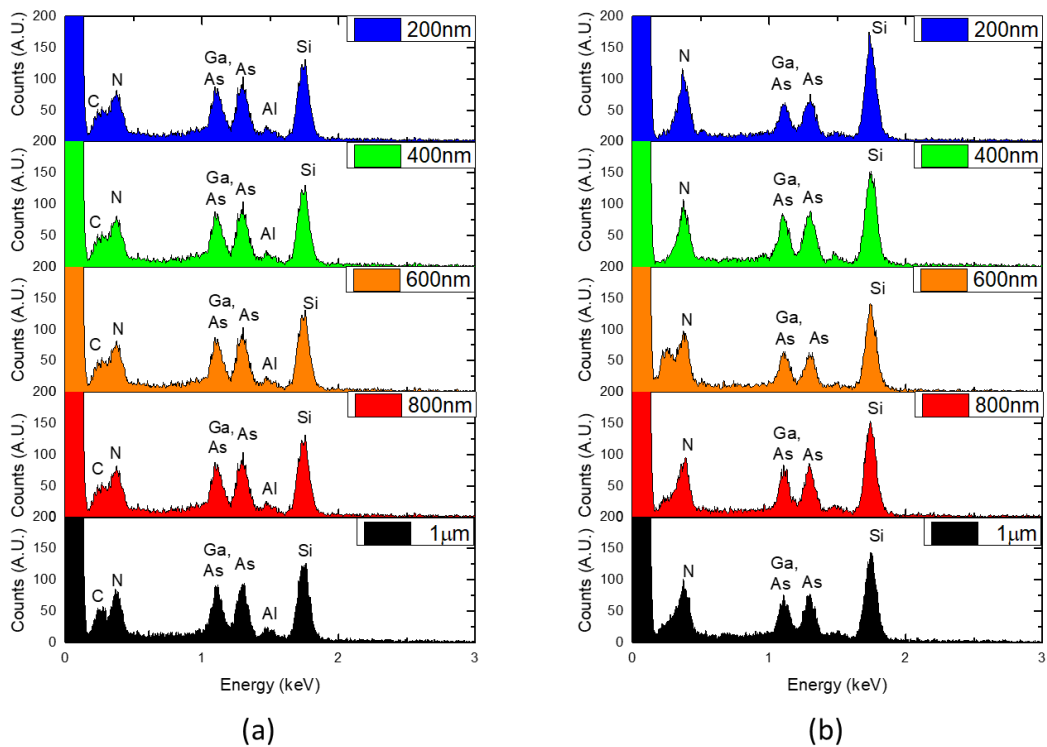

**Figure S4:** EDS spectra for nanopillars varying with size ranging from 200 nm-1 μm. (a) Sample coated with LF-PECVD $Si_xN_y$ without sulfurization. (b) Sample pre-treated with ammonium sulfide and coated with LF-PECVD $Si_xN_y$. (In all measurements a voltage of 5 kV was used).

## S7. X-ray photoelectron spectroscopy (XPS)

*Experiment*

For further explaining the potential of the LF-PECVD $Si_xN_y$ to passivate GaAs, additional measurements using X-ray photoelectron spectra (XPS) were performed. Considering the typical analyzed large area covered by our XPS ssytem (650 μm × 650 μm), for the measurements additional samples of ~500 nm etched GaAs (without patterned pillars) were prepared, following the same etching procedure conditions (see S1) as performed for the nanopillars. The remaining passivation treatments followed identical conditions as discussed earlier (section S2) and here we focused only on samples using the $Si_xN_y$-based best treatments: (a) unpassivated sample, (b) sample with LF-PECVD $Si_xN_y$ coating only (treatment #6), (c) sample with ammonium sulfide passivation and coating of $Si_xN_y$ by LF-PECVD (treatment #5), (d) sample with ammonium sulfide passivation followed by coating using HF-PECVD $Si_xN_y$ (treatment #4). The key difference of the treatments employed is that since XPS spectra can effectively measure only thicknesses within 10 nm from the surface, for these samples a thickness of only ~4 nm of $Si_xN_y$ (instead of ~80 nm) was deposited coating the etched GaAs surface. The XPS spectra was collected using an ESCALAB 250Xi system (Thermo Scientific) in UHV ($< 10^{-9}$ Torr). The monochromatic Al-Kα source (1486.6 eV) was used to analyze an area of 650 μm × 650 μm in the prepared samples. For fitting the peaks of measured spectra we used Avantage software with Voigt functions (convolution of Lorentzian and Gaussian functions). As discussed next the peak fittings closely match the standard chemical states previously recorded in literature.[1,2] The calibration of the peaks was done by adventitious carbon peak (284.8 eV).

*Results and discussion*

Unlike other semiconductors, as for example silicon, the native oxides formed at the surface of GaAs are not stable. This contributes to non-radiative recombination sites at the GaAs surface limiting the PL of the GaAs material.[3] Through passivation methods, the removal of these oxide layers and respective replacement with stable coating material is achieved. Thus the quality of the passivation treatment can be quantified by the minimum amount of surface defects formed by the native oxides of GaAs (here Ga-O ($Ga_2O_3$) and As-O ($As^{3+}$ and $As^{5+}$)). Here we focus our analysis on the removal of these native oxides using the best treatments shown in PL measurements employing $Si_xN_y$ coating layers.

Figure S5 (identical figure as Fig. 4 in the main paper) shows the Ga 3d XPS spectra comparison for an untreated sample, Fig. S5(a), and for samples using various $Si_xN_y$-based surface treatments, Figs. S5(b)-(d). First we analyze the passivation using LF-PECVD $Si_xN_y$ without any sulfurization pre-treatment. In the unpassivated case, spectrum of Figure S5(a), we observe a high energy shoulder which is less pronounced for the LF-PECVD $Si_xN_y$ treatment, Figure S5(b). This indicates suppression of the Ga native oxide (Ga-O) peak (binding energy (B.E.) ~20 eV, blue trace), indicating the treatment with LF-PECVD without pre-treatment provides already an impact on the removal of gallium oxides. Noteworthy, this effect is already noticeable even in the case of a thin deposited layer (~4 nm). We note this thin layer was a requirement in our experiments to be able to perform the XPS analysis.

Next we compare the LF-PECVD $Si_xN_y$ treatment versus the HF-PECVD $Si_xN_y$, Figs. S5(c) and (d), respectively. In both cases an ammonium sulfide pre-treatment was used. Clearly in both cases the GaAs peak (binding energy ~19.2 eV) is the prominent peak whereas Ga native oxides (Ga-O) are insignificant. This shows the success of combining the ammonium sulfide and $Si_xN_y$ coatings for the removal of native oxides. When comparing in more detail both cases, we observe a broader and larger Ga-O peaks for the HF-PECVD $Si_xN_y$ coated sample, panel (d), as compared to the LF-PECVD $Si_xN_y$ coated sample, panel c). This indicates a better performance of the LF-PECVD $Si_xN_y$ cases as compared with the HF-PECVD $Si_xN_y$. The results are confirmed in Table S1 which summarizes the ratio of the atomic percentage (at %) of Ga-O to GaAs. A low at % ratio (~ 0.14) is achieved for LF-PECVD $Si_xN_y$ coated sample which indicates the least presence of Ga-O for the best treatment and in line with the trend observed in PL measurements.

**Table S1:** Fitting parameters and atomic percentage ratio of the XPS spectra of Ga 3d.

| Treatment | GaAs B.E. (eV) | GaAs Peak FWHM (eV) | Ga-O B.E. (eV) | Ga-O Peak FWHM (eV) | Ratio of at % (Ga-O /GaAs) |
|---|---|---|---|---|---|
| Unpassivated | 19.55 | 1.07 | 20.65 | 1.45 | 0.88 |
| LF-PECVD $Si_xN_y$ only | 19.69 | 1.1 | 20.59 | 1.36 | 0.78 |
| $(NH_4)_2S$ + LF-PECVD $Si_xN_y$ | 19.21 | 1.06 | 19.99 | 1.28 | 0.14 |
| $(NH_4)_2S$ + HF-PECVD $Si_xN_y$ | 19.16 | 1.03 | 19.81 | 1.53 | 0.27 |

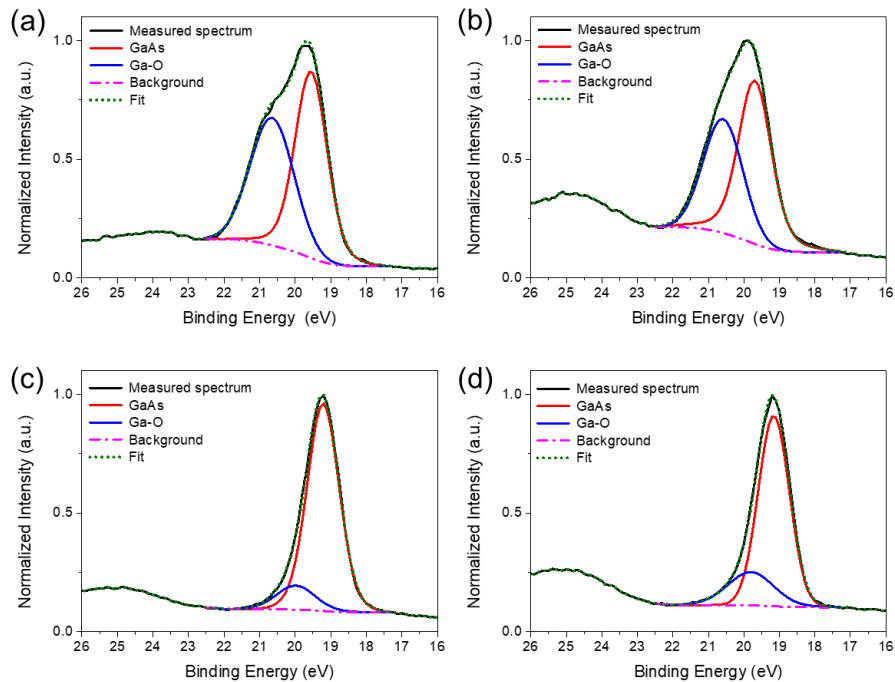

**Figure S5:** XPS spectra of Ga 3d for: (a) Unpassivated sample. (b) Sample coated using LF-PECVD $Si_xN_y$. (c) Sample using ammonium sulfide treatment followed by LF-PECVD $Si_xN_y$ coating. (d) Sample using ammonium sulfide treatment followed by HF-PECVD $Si_xN_y$ coating.

[The results in this figure are also shown in the main paper, Fig. 4.]

In what follows we analyze the As 3d XPS spectra, Fig. S6. First we compare the unpassivated sample, panel (a), with a sample using LF- PECVD $Si_xN_y$ treatment without sulfurization, panel (b). A shift in As-O peaks (both $As^{3+}$ and $As^{5+}$) is observed for the sample with LF-PECVD $Si_xN_y$ treatment as compared with the unpassivated sample. Further, a prominence in $As^{5+}$ peaks ($As^{5+}$ $3d_{5/2}$ and $3d_{3/2}$) is observed for the sample with LF-PECVD $Si_xN_y$ treatment, panel (b), as compared to the prominent $As^{3+}$ peaks ($As^{5+}$ $3d_{5/2}$ and $3d_{3/2}$) for the untreated sample, panel (a). This indicates an effect of the LF-PECVD $Si_xN_y$ treatment on the GaAs surface. Possibly the $Si_xN_y$ film additionally participates directly in the formation of interfacial bonding at the GaAs surface.

Table S2 shows the As 3d fitting parameters and the at % ratio of As-O to GaAs. The at % ratio is 1.27 for the passivated sample, which compares with an at % ratio of 1.52 for the unpassivated sample. The lower ratio indicates oxide removal is achieved despite the very thin layer deposited. Noteworthy the samples treated with ammonium sulfide followed by the $Si_xN_y$ deposition, Figs. S6 (c), (d), show a clear suppression of native oxides. Here, due to the successful suppression of As-O peaks in both cases the effect of the plasma frequency (LF vs HF $Si_xN_y$ PECVD) was not possible to compare.

**Table S2:** Fitting parameters and atomic percentage ratio of the XPS spectra of As 3d.

| Treatment | GaAs $3d_{5/2}$ B.E., FWHM (eV) | GaAs $3d_{3/2}$ B.E., FWHM (eV) | As-O ($As^{3+}$) $3d_{5/2}$ B.E., FWHM (eV) | As-O ($As^{3+}$) $3d_{3/2}$ B.E., FWHM (eV) | As-O ($As^{5+}$) $3d_{5/2}$ B.E., FWHM (eV) | As-O ($As^{5+}$) $3d_{3/2}$ B.E., FWHM (eV) | Ratio of at % (As-O/ GaAs) |
|---|---|---|---|---|---|---|---|
| Unpassivated | 41.31, 0.99 | 42.03, 0.99 | 44.45, 1.63 | 45.2, 1.63 | 45.78, 1.29 | 46.54, 1.29 | 1.52 |
| LF PECVD $Si_xN_y$ only | 41.48, 1.06 | 42.19, 1.06 | 44.26, 1.42 | 44.89, 1.42 | 45.3, 1.49 | 46.1, 1.49 | 1.27 |
| $(NH_4)_2S$ + LF PECVD $Si_xN_y$ | 40.91, 0.95 | 41.61, 0.95 | N/A | N/A | N/A | N/A | N/A |
| $(NH_4)_2S$ + HF PECVD $Si_xN_y$ | 40.85, 0.92 | 41.55, 0.92 | N/A | N/A | N/A | N/A | N/A |

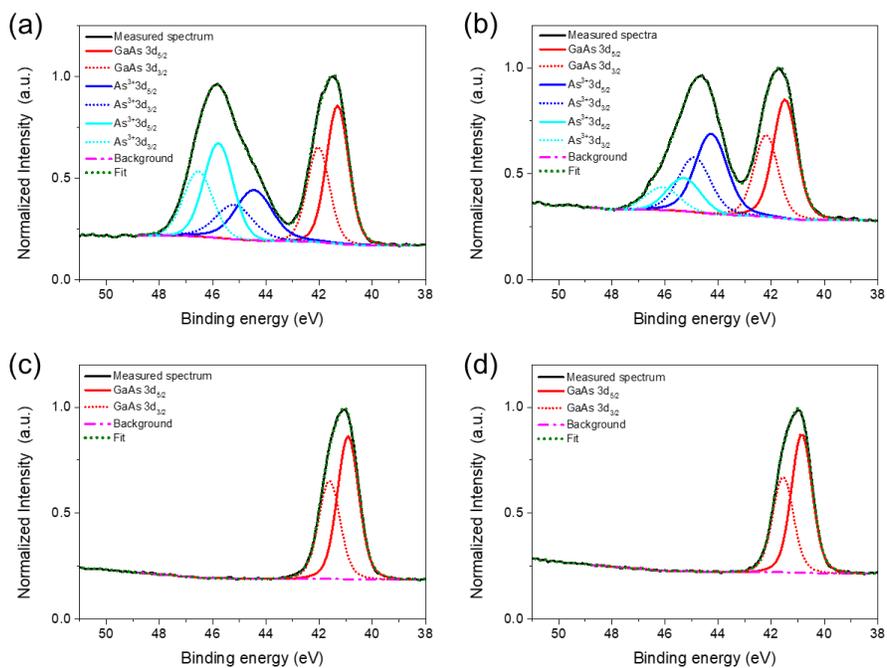

**Figure S6:** XPS spectra of As 3d for: (a) Unpassivated sample. (b) Sample coated using LF-PECVD $Si_xN_y$. (c) Sample using ammonium sulfide treatment followed by LF-PECVD $Si_xN_y$ coating. (d) Sample using ammonium sulfide treatment followed by HF-PECVD $Si_xN_y$ coating.